\documentclass[journal]{IEEEtran}
\usepackage{stackengine}
\usepackage{amssymb}
\usepackage{amsmath}
\usepackage{dsfont}

\usepackage{amssymb}
\usepackage{cite}
\usepackage{graphicx}
\usepackage{bm}
\usepackage{multirow}
\usepackage{booktabs}
\usepackage{color}
\usepackage{upgreek}
\usepackage[bookmarks={false}]{hyperref}
\usepackage[hyphenbreaks]{breakurl}
\usepackage{balance}
\usepackage{verbatim}
\usepackage{relsize}
\usepackage{nomencl}
\usepackage{ifthen}
\usepackage{soul}

\usepackage{accents}
\usepackage{setspace}
\usepackage{bbold}
\usepackage[compact]{titlesec}
\usepackage{tikz}
\usepackage{amssymb}
\usepackage{cuted}
\usepackage{float}
\usepackage{nicematrix}
\usepackage[linesnumbered,ruled,vlined]{algorithm2e}
\usepackage{makecell}
\makeatletter
\renewcommand\subsubsection{\@startsection{subsubsection}{3}{\parindent}
  {0.1\baselineskip} 
  {0.1\baselineskip} 
  {\normalfont\normalsize\itshape}} 
\makeatother
\begin{document}
\title{Model Predictive Black Start for Dynamic Formation of DER-Led Microgrids with Inrush Current Impacts}

\author{Cong~Bai,~\IEEEmembership{Student~Member,~IEEE,} Salish~Maharjan,~\IEEEmembership{Member,~IEEE,} and~Zhaoyu~Wang,~\IEEEmembership{Senior~Member,~IEEE}
\vspace{-1cm}}
\maketitle
\begin{abstract}
Black start (BS) of the distribution system (DS) with high penetration of distributed energy resources (DERs) require advanced control frameworks to ensure secure and efficient restoration. This paper proposes a model predictive black start (MPBS) framework incorporating an inrush current feasibility module to dynamically generate real-time feasible and optimal restoration sequences. Short-term forecasts of DER output and transmission grid (TG) availability are utilized to construct adaptive cranking paths. The inrush current feasibility module analytically estimates the transient inrush current caused by energizing no-load distribution transformers (DTs). To mitigate excessive inrush current and avoid potential misoperations of protection devices, an emergency operation-inspired voltage control strategy and a switch blocking mechanism are developed. The proposed inrush model is validated against electromagnetic transient (EMT) simulations in PowerFactory with estimation accuracies exceeding $90\%$. Case studies on a modified IEEE 123-node feeder demonstrate that the MPBS framework prevents misoperations of fuses and reclosers, reduces unnecessary DER energy consumption, and enhances load restoration efficiency during DER-led BS processes.
\end{abstract}
\IEEEpeerreviewmaketitle
\section{Introduction}
\IEEEPARstart{T}{he} increasing penetration of distributed energy resources (DERs) in distribution systems (DSs) is fundamentally reshaping traditional power system restoration practices~\cite{O’Brien2022}. In the face of large-scale disruptions, such as extreme weather events or severe cyberattacks, it has become essential for DSs to possess the capability to rapidly and securely self-restore following the loss of connection to the upstream transmission grid (TG)~\cite{Ding2022}. As centralized diesel or thermal generation becomes less accessible or delayed, black start (BS) operations in the DS are increasingly driven by DERs, particularly grid-forming inverter (GFMI)-based battery energy storage systems (BESSs) and behind-the-meter (BTM) photovoltaics (PVs)~\cite{Wang2015}. These resources enable the dynamic formation of microgrids (MGs) and support localized, resilient restoration.

Conventional BS strategies for DER-dominated DSs often rely on static cranking paths generated by rule-based controller, which are insufficiently responsive to time-varing DER outputs, dynamic load conditions, and changing TG availability~\cite{Poude2019}. Furthermore, when BS is initiated from multiple sources, such as dynamically formed MGs, the restoration process must address complex challenges, including protection coordination, voltage regulation, and staged load recovery~\cite{Banerjee2023}. A critical concern is the surge of inrush current associated with the energization of no-load distribution transformers (DTs), which can lead to the misoperation of protection devices like fuses and reclosers~\cite{Shahparasti2022}. These protection misoperations can interrupt the restoration sequence, damage equipment, or prolong the outage duration.

Recent works have explored model predictive control (MPC) for adaptive BS and MG coordination, leveraging real-time measurements to improve restoration efficiency. In particular, MPC-based generator start-up optimization strategies have been introduced to dynamically determine optimal cranking sequences, where frequency deviations of distributed MGs are constrained~\cite{Zhao2018}. To manage more complex system architectures, hierarchical MPC structures have been proposed to coordinate wind farms and BESS during BS, where frequencies are regulated at both local and system levels~\cite{Liu2020}. Similarly, fuzzy MPC techniques have been developed for temporary MGs, enabling robust frequency regulation under model uncertainty and DER variability~\cite{Wang2023}.

Recognizing the increasingly decentralized nature of restoration, researchers have introduced double-layer and real-time conditional MPC frameworks tailored for interconnected MGs. These frameworks support decentralized self-restoration actions while maintaining overall system coherence and avoiding inter-MG conflicts during the BS process~\cite{Hu2021}. Furthermore, simulation-assisted restoration models and collaborative DS restoration strategies have emerged to integrate both pre-event planning and real-time dispatch layers. These approaches consider complex factors such as BTM DER participation, voltage constraints, and real-time frequency dynamics to improve restoration precision and operational realism~\cite{Liu2021,Zhang2021}. A recent study formulated a real-time MPC-based BS method for DER-dominated DSs, which efficiently synchronizes MGs and sequences restoration actions based on evolving system measurements~\cite{Konar2023}. However, most of these advanced MPC-based frameworks prioritize voltage and frequency regulation while overlooking the protection-related challenges—especially the risks associated with transient inrush currents that can arise when energizing no-load DTs. These surges can lead to the unintended operation of protection devices, potentially interrupting the restoration process. As a result, while existing literature has significantly advanced the coordination and optimization of BS actions, the omission of inrush current modeling and its implications on protection coordination leaves an essential gap for practical deployment in DSs.

To bridge this critical gap, recent studies have begun to examine the impact of inrush current during BS and propose mitigation strategies to ensure protection coordination in DER-rich DSs. Simulation tools such as dynamic phasor modeling have been used to analyze inrush current in unbalanced distribution networks, revealing its nonlinear and time-varying characteristics that can cause protection miscoordination and hinder restoration continuity~\cite{Elizondo2017}. Traditional mitigation strategies rely on transformer-side solutions, such as controlled switching or additional damping components, but these methods require extra hardware and cost~\cite{Alassi2023}. To address this, recent studies have incorporated predictive control schemes at the inverter level to proactively suppress inrush current. For instance, a finite control set MPC combined with an autoregressive model was proposed to predict and reduce magnetizing inrush during BS, achieving robust mitigation without detailed transformer modeling~\cite{Bainian2025}. These advances highlight the potential of embedding inrush feasibility checks directly within restoration controllers, yet they remain largely confined to component-level validation without integration into broader DS-wide restoration. Thus, a system-level MPC framework that considers inrush constraints for network-wide DER-led BS coordination remains an open and essential research direction.

Hence, this paper proposes a model predictive black start (MPBS) framework integrated with an inrush current feasibility module to address the risk of protection device misoperations during the restoration process. The inrush current caused by energizing no-load DTs is estimated and mitigated within the BS sequence. The main contributions of this work are summarized as follows:
\begin{itemize}
    \item A MPBS framework is proposed to adaptively coordinate DER-led restoration in the DS. The framework updates real-time DER output, load demand, and TG availability over a prediction horizon, generating an optimal cranking path through receding-horizon control.
    \item An analytical inrush current model is developed for the energization of no-load DTs during the BS. The model estimates current surges based on residual flux and switching angle, and serves as a feasibility check within the control loop. Accuracy is validated against electromagnetic transient (EMT) simulations in PowerFactory.
    \item An emergency operation-inspired voltage reduction strategy is integrated into a ZIP load model to mitigate inrush-induced protection misoperations. By actively reducing GFMI voltage setpoints, the strategy suppresses transient demand spikes, enhancing system security and energy efficiency during restoration.
\end{itemize}

The remainder of this paper is organized as follows. Section~\ref{sec:ii} presents the overall structure of the proposed MPBS framework. Section~\ref{sec:iii} introduce the inrush current feasibility module in detail. Section~\ref{sec:iv} describes the mathematical formulation of the restoration problem considering inrush current impacts. Section~\ref{sec:v} designs three case studies to validate the effectiveness of the developed framework. Finally, Section~\ref{sec:vi} concludes the entire work.
\section{Overview of the Proposed Model Predictive Black Start (MPBS) Framework}\label{sec:ii}
\begin{figure}[htbp]
    \centering
    \includegraphics[width=0.9\linewidth]{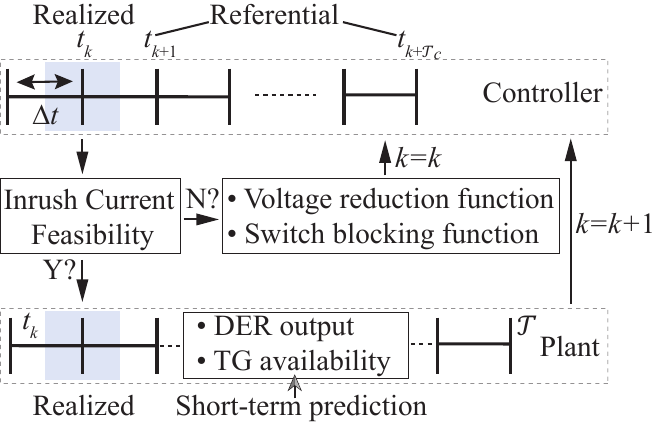}
    \vspace{-1em}
    \caption{Inrush current equivalent process.}
    \label{fig:MPBS}
\end{figure}

The overall structure of the proposed model predictive black start (MPBS) framework is illustrated in Fig.~\ref{fig:MPBS}. It consists of three main components: the controller, the inrush current feasibility module, and the plant. After a blackout event, the DS is disconnected from the TG, and distributed GFMIs are deployed as BS resources. These GFMIs independently form MGs across the DS and begin the restoration process in parallel. Synchronization across multiple MGs is enabled through SSWs, allowing them to flexibly expand their boundaries as the restoration progresses~\cite{Maharjan2025}.

At each time step $t_{k}$ within a predefined restoration horizon $\mathcal{T}$, the plant performs short-term prediction of DER output and TG availability over the controller's prediction window $\mathcal{T}_c$. This information is sent to the controller, which then solves an optimization problem to generate a restoration cranking path over the same horizon. Only the first step of the generated path is intended for immediate execution in the plant, while the remaining steps serve as a reference for future decisions. Before realization, the proposed path is passed to the inrush current feasibility module, which estimates the inrush current caused by planned switch closures and evaluates whether the resulting current may cause misoperation of protection devices such as fuses or reclosers.

If the restoration cranking path is deemed feasible, the first-step control action is implemented in the plant. Otherwise, inrush mitigation strategies are activated. These include: (1) a voltage reduction function inspired by emergency operation, which reduces load demand and inrush severity by lowering the output voltage of the GFMI; and (2) a switch blocking function that excludes high-risk switching actions from the feasible space. The modified conditions are fed back to the controller, which re-solves the optimization problem at the current time step $t_{k}$. This feedback loop ensures that restoration actions remain both optimal and feasible in real time. The overall process is executed repeatedly across the restoration horizon $\mathcal{T}$, progressively restoring more load until system recovery is maximized.

To support the MPBS framework, the DS is formally represented as a graph $\Gamma = (\mathcal{B},\mathcal{E}, \Phi)$, where $\mathcal{B}$, $\mathcal{E}$, and $\Phi$ denote the sets of buses, edges (or branches), and phases, respectively. The DS is partitioned into multiple segments, referred to as bus blocks, which serve as the basic units of restoration. These bus blocks are interconnected via switches, including energizing switches (ESWs) and synchronizing switches (SSWs), and are indexed by the set $\mathcal{G}$. A set $\mathcal{L}$ is used to represent the laterals in the DS, which typically originate from main feeders and supply power to downstream loads. A subset of buses, denoted $\mathcal{B}^{\mathrm{GFMI}}$, is selected as candidate locations for GFMI deployment due to their self-starting capability. Based on the installed GFMIs, a set of microgrids $\mathcal{M}$ is formed, each acting as an autonomous BS source responsible for initiating and coordinating the local restoration process.
\section{Modeling of Inrush Current Feasibility}\label{sec:iii}
In this section, the inrush current resulting from energizing a no-load distribution transformer during the BS is first estimated using the source-side node voltage and the dynamic nodal impedance matrix. Subsequently, the inrush currents are evaluated on a per-lateral and per-microgrid basis to assess the risk of protection device misoperation. Finally, two mitigation strategies, including voltage reduction and switch blocking, are introduced to address infeasible inrush currents.
\subsection{Estimation of Inrush Current during Black Start (BS)}
BS involves the sequential switching of de-energized bus blocks. Energizing a dead bus block typically induces a large transient current, commonly referred to as the inrush current. In the case of a single-phase no-load DT, if the initial phase angle of the applied voltage on the high-voltage side, denoted as $\theta$, aligns with the residual flux $\lambda_0$, the core may enter saturation and produces a very large magnetizing current, which is one of the primary contributors to the inrush current. Given the nominal flux and positive saturation flux of the DT, denoted by $\lambda_n$ and $\lambda_s$, respectively, a saturation indicator function $h(\theta)$ is defined to characterize the magnetic core status during energization:
\begin{equation}\label{eq:saturationindicator}
    h(\theta) = \begin{cases}
    1,& \mathrm{if}\quad\lambda_n\cos{\theta} > \lambda_s - \lambda_n - \lambda_0; \\
	-1, & \mathrm{if}\quad\lambda_n\cos{\theta} < \lambda_n - \lambda_s - \lambda_0; \\
    0, & \mathrm{otherwise}, \\
	\end{cases}
\end{equation}
where $\pm1$ represent that the core enters saturation in the positive and negative flux directions, respectively, while $0$ means that the core remains unsaturated.

Assume the $\lambda_0$ is known and fixed prior to energizing the DT, and $h(\theta)\in\{-1,1\}$, the peak inrush current $i_{pk}$ for a specific $\theta$ can be estimated as~\cite{Jazebi2015}:
\begin{equation}\label{eq:singleinrushcurrent}
    i_{pk} = \left(\frac{\lambda_0 - h(\theta)\lambda_s}{\lambda_n} + \cos{\theta} + 1\right)i_{\mathrm{SS}},
\end{equation}
where $i_{\mathrm{SS}}$ denotes the magnitude of steady-state part of the saturated current on the high-voltage side of the DT.

\begin{figure}[htbp]
    \centering
    \includegraphics[width=\linewidth]{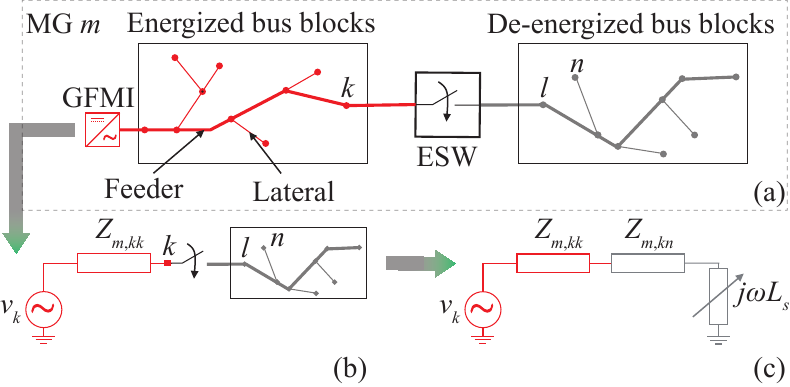}
    \vspace{-2em}
    \caption{Inrush current equivalent process. (a) MG while energizing a bus block. (b) Thevenin's equivalent. (c) Mutual impedance and saturated reactance equivalent.}
    \label{fig:inrush_current_equi_ckt}
\end{figure}

To calculate the $i_{\mathrm{SS}}$, it is necessary to obtain the dynamic nodal impedance matrix of the currently formed MG. For each $m\in\mathcal{M}$, let $Z_{m}$ denote the nodal impedance matrix of MG $m$, which evolves dynamically as the MG boundary expands, both its dimensions and elements change in response to the energized topology. To estimate $i_{\mathrm{SS}}$ during the closing of an ESW across $(k,l)$, we adopt the equivalent DT model illustrated in Fig.~\ref{fig:inrush_current_equi_ckt}(a)-(c). In this model, the energized portion of MG $m$ is represented by its Thevenin's equivalent, while the energizing segment between the energized portion and the saturated core of the DT is modeled using the mutual impedance, $Z_{m,kn}$, between the source-side node $k$ and the pick-up node $n$. The Thevenin voltage corresponds to the measured voltage, $v_k$, at bus $k$, and the Thevenin impedance is approximated by the diagonal element $Z_{m,kk}$ of the matrix $Z_{m}$. The reactance of the DT's saturated core is given by $\omega L_s$, where $\omega$ is the angular frequency and $L_{s}$ is the saturated inductance.
Hence, the $i_{\mathrm{SS}}$ is estimated as follows:
\begin{equation}\label{eq:steadystatesaturatedcurrent}
    i_{\mathrm{SS}} = \frac{v_k}{|Z_{m,kk} + Z_{m,kn} + j\omega L_s|}
\end{equation}
where $|\cdot|$ is the operator used to calculate the absolute value of a complex number.
\subsection{Feasibility check of Inrush Current during BS}
The inrush current resulting from energizing a no-load DT may interfere with existing protection devices, such as fuses and reclosers. Therefore, the restoration cranking path generated by the controller must be validated for inrush current feasibility before implementing in the plant.

Assume that all DTs on the same phase retain identical residual flux values from the moment of blackout to the beginning of the BS, the $\theta$ that produces the maximum inrush current for a specific phase during the energizing of a bus block can be determined with Eq.~\eqref{eq:singleinrushcurrent}. Furthermore, by accounting for the distribution of DTs across different phases within a given bus block $g$, the worst-case (WC) closing phase angle $\theta_{\mathrm{WC},g}$, referenced to phase A, can be identified as the angle that yields the highest inrush current across the entire bus block prior to restoration. Subsequently, the impact of this worst-case inrush current on protection devices can be evaluated and expressed as follows.
\subsubsection{Impact on Fuses}
A fuse is installed at the head of each lateral to protect against overcurrent across the entire downstream zone. During the energization of a bus block, the total inrush current at the head of a lateral is calculated as the sum of the inrush current contributed by the saturated DTs within that lateral. This can be expressed as:
\begin{equation}
    \boldsymbol{i}_{pk,l}^{\mathrm{L}} = \sum_{b\in\mathcal{B}_{l}^{\mathrm{L}}}\boldsymbol{i}_{pk,b}^{\mathrm{DT}}, \forall l\in\mathcal{L}, 
\end{equation}
where $\boldsymbol{i}_{pk,l}^{\mathrm{L}}$ and $\boldsymbol{i}_{pk,b}^{\mathrm{DT}}$ are column vectors representing the inrush current contributions associated with lateral $l$ and DT $b$, respectively.

Let $\boldsymbol{i}_{2\mathrm{T},l}^{mel}$ denote the vector of minimum melting currents for two cycles of the fuses on lateral $l$, the fuse status during the energizing of the corresponding bus block can be evaluated using the following indicator function:
\begin{equation}
    \boldsymbol{y}_{l}^{\mathrm{F}} = \mathbb{1}\left(\boldsymbol{i}_{pk,l}^{\mathrm{L}}>\boldsymbol{i}_{2\mathrm{T},l}^{mel}\right),\forall l\in\mathcal{L},
\end{equation}
where $\boldsymbol{y}_{l}^{\mathrm{F}}\in\{0,1\}^{|\Phi_l|}$ is a column indicator vector that flags whether the fuse on each phase of lateral $l$ has melted, and $\mathbb{1}(\cdot)$ denotes an element-wise indicator function. The term $|\Phi_l|$ represents the cardinality of the set $\Phi_l$, which stores the phases of the lateral $l$.
\subsubsection{Impact on Reclosers}
A recloser, typically installed at the head of the main feeder, serves as a short-time delay overcurrent protection device and is capable of multiple operations in response to transient events such as inrush current. When a GFMI is used as a BS source to form an MG and restore load, the recloser located near it must withstand the total inrush currents resulting from the energization of multiple bus blocks. Let $\boldsymbol{i}_{pk,m}^{\mathrm{MG}}$ denote the total peak inrush currents flowing through the recloser at the feeder head of MG $m$. It can be expressed as:
\begin{equation}
    \boldsymbol{i}_{pk,m}^{\mathrm{MG}} = \sum_{g\in\mathcal{G}_{m}}\sum_{l\in\mathcal{L}_{g}}\boldsymbol{i}_{pk,l}^{\mathrm{L}},\forall m\in\mathcal{M},
\end{equation}
where $\mathcal{L}_g$ denotes the set of laterals associated with bus block $g$, and $\mathcal{G}_m$ represents the set of bus blocks contained within the MG $m$.

Let $\boldsymbol{i}_{2\mathrm{T},m}^{act}$ denote the vector of rapid act currents for two cycles of the recloser belonging to the MG $m$, the recloser status during the expansion of MG can be evaluated using the following indicator function:
\begin{equation}
    \boldsymbol{y}_{m}^{\mathrm{R}} = \mathbb{1}\left(\boldsymbol{i}_{pk,m}^{\mathrm{MG}}>\boldsymbol{i}_{2\mathrm{T},m}^{act}\right),\forall m\in\mathcal{M},
\end{equation}
where $\boldsymbol{y}_{m}^{\mathrm{R}}\in\{0,1\}^{|\mathcal{M}|}$ is a column indicator vector that flags whether each recloser has opened.
\subsection{Mitigation of Inrush Current during BS}
Effective strategies are required to mitigate infeasible inrush currents during the BS process, ensuring that the generated restoration cranking path does not unintentionally trigger protection devices.
\subsubsection{Voltage Reduction Function}
A fuse located at the head of a lateral may blow if the downstream inrush current, resulting from the energization of no-load DTs, exceeds its minimum melting current threshold. Considering the WC closing phase angle, the most effective way to mitigate the inrush current through the fuse is to reduce the source-side node voltage, as described by Eq.\eqref{eq:steadystatesaturatedcurrent}.

Inspired by the emergency operation technique~\cite{ANSI_C84.1_2020}, an additional voltage reduction mechanism is introduced for each bus $b\in\mathcal{B}^{\mathrm{GFMI}}$, where the voltage is regulated by the GFMI. This reduction is conditionally applied based on the fuse's status within the corresponding MG $m$, and is formulated as:
\begin{subequations}\label{eq:gfmivoltage}
\begin{align}
    \boldsymbol{v}_{b,t} \le &\lceil \boldsymbol{v}_{b} \rceil + \mathop{\textstyle\bigcirc}_{l\in\mathcal{L}_{m}}\left(\boldsymbol{y}_{l}^{\mathrm{F}}\right)\circ\left(\boldsymbol{v}^{red} - \lceil \boldsymbol{v}_{b} \rceil\right)\label{eq:gfmivoltage_r},\\
    \boldsymbol{v}_{b,t} \ge &\lfloor \boldsymbol{v}_{b} \rfloor + \mathop{\textstyle\bigcirc}_{l\in\mathcal{L}_{m}}\left(\boldsymbol{y}_{l}^{\mathrm{F}}\right)\circ\left(\boldsymbol{v}^{red} - \lfloor \boldsymbol{v}_{b} \rfloor\right)\label{eq:gfmivoltage_l},
\end{align}
\end{subequations}
where $\boldsymbol{v}_{b,t}$ is a column vector representing the squared voltage magnitude at time $t$, the symbols $\lfloor \cdot \rfloor$ and $\lceil \cdot \rceil$ denote the lower and upper bound operators of a variable, respectively, '$\circ$' represents the Hadarmard product between two vectors of the same dimension, $\bigcirc(\cdot)$ denotes the Hadamard product applied across a set of vectors, and $\boldsymbol{v}^{red}$ is a predefined column vector specifying the reduced voltage levels to be enforced by the GFMI when the fuse operation is detected within the associated MG by the inrush current feasibility check module.
\subsubsection{Switch Blocking Function}
To ensure safe operation during the BS process, the reduced voltage applied by the voltage reduction function must not be excessively low. When a large number of bus blocks are energized by the MG, voltage reduction alone may be insufficient to effectively manage the resulting inrush current. To address this limitation, a switch blocking function is introduced as a complementary mitigation strategy.

For the MG $m$ formed within the restoration cranking path determined by the controller in Fig.~\ref{fig:MPBS}, if the associated recloser is triggered at the realized step, the optimization problem is resolved at the time step $t$ with the following constraint added:
\begin{equation}\label{eq:switchblockingfunction}
    \left\|\boldsymbol{y}^{\mathrm{R}}_{m}\right\|_{\infty}u^{\mathrm{ESW}}_{k^*n^*,t} = 0,
\end{equation}
where $u^{\mathrm{ESW}}_{k^*n^*,t}$ is a binary variable indicating whether the ESW $(k^*,n^*)$ is closed at time $t$. The ESW $(k^*,n^*)$  is selected based on the following criterion:
\begin{equation}
    (k^*,n^*) = \underset{\substack{(k,n)\in\mathcal{E}^{\mathrm{ESW}}_{g}\\g\in\mathcal{G}_m}}{\arg\max} \; \sum_{l\in\mathcal{L}_{g}}\left\|\boldsymbol{i}_{pk,l}^{\mathrm{L}}\right\|_1,
\end{equation}
where $\mathcal{E}^{\mathrm{ESW}}_{g}$ denotes the set of ESWs associated with bus block $g$. The constraint blocks the closure of the ESW that would lead to the highest peak inrush current across all laterals in the bus block associated with the recloser's activation.
\section{Mathematical Formulation of The BS Considering Inrush Current Feasibility}\label{sec:iv}
This section presents the mathematical formulation of BS problem, explicitly incorporating the inrush current feasibility. First, the objective function is defined, capturing the weighted restoration of different types of loads. Next, the dynamic energization process of the DS is modeled, enabling the formation and expansion of MGs throughout the restoration horizon. Then, the operational constraints governing the power output of various devices and their relationship are described. Finally, the complete solution structure of the proposed MPBS framework is summarized in a compact mathematical form.
\subsection{Objective of BS}
The objective is to maximize the total restored loads over the restoration horizon. This is formulated as:
\begin{equation}\label{eq:objecivefunction}
    \max\sum_{t\in\mathcal{T}_c}\Delta t\sum_{\mathcal{X}\in\{\mathrm{CL}, \mathrm{NL}\}}\gamma^{\mathcal{X}}\sum_{b\in\mathcal{B}^{\mathcal{X}}} \left(\boldsymbol{1}^{T}_{|\Phi_b|}\boldsymbol{p}^{\mathcal{X}}_{b,t}\right),
\end{equation}
where $\Delta t$ is the time step duration, and $\mathcal{X}\in\{\mathrm{CL}, \mathrm{NL}\}$ denotes the load types, including critical load (CL) and non-critical load (NL). The coefficient $\gamma^{\mathcal{X}}$ represents the weight assigned to load type $\mathcal{X}$, while $\mathcal{B}^\mathcal{X}$ is the set of buses with loads of type $\mathcal{X}$. The set $\Phi_b$ contains the phases of loads connected to bus $b$, and $\boldsymbol{1}_{|\Phi_b|}^{T}$ is the transpose of an all-ones column vector of dimension $|\Phi_b|$, used to sum the phase-wise active load demand, $\boldsymbol{p}^{\mathcal{X}}_{b,t}$.
\subsection{Formulation of BS Energization}
\subsubsection{Bus Block Energization Status}
Given the set of non-switchable lines in the DS as $\mathcal{E}^{\mathrm{LN}}$, and the set of edges within a bus block $g\in\mathcal{G}$ as $\mathcal{E}_g$, the subset of non-switchable lines contained in bus block $g$ is defined as $\mathcal{E}^{\mathrm{LN}}_g = \mathcal{E}^{\mathrm{LN}} \cap \mathcal{E}_g$. Similarly, the subset of ESWs within bus block $g$ is given by $\mathcal{E}^{\mathrm{ESW}}_g = \mathcal{E}^{\mathrm{ESW}} \cap \mathcal{E}_g$. 

The energization status of each bus block $g$ is constrained by the following conditions:
\begin{subequations}\label{eq:segmentstatus}
    \begin{align}
    u^{\mathrm{BB}}_{g,t} &\ge u^{\mathrm{BB}}_{g,t - 1},\label{eq:segmentstatus1}\\
    u^{\mathrm{BB}}_{g,t} &= u^{\mathrm{B}}_{b,t}, \forall b\in\mathcal{B}_g,\label{eq:segmentstatus2}\\
    u^{\mathrm{BB}}_{g,t} &= u^{\mathrm{E}}_{kl,t}, \forall (k,l)\in\mathcal{E}^{\mathrm{LN}}_g,\label{eq:segmentstatus3}\\
    u^{\mathrm{BB}}_{g,t} &\ge u^{\mathrm{ESW}}_{kl,t}, \forall (k,l)\in\mathcal{E}^{\mathrm{ESW}}_g,\label{eq:segmentstatus4}\\
    \textstyle\sum_{(k,l)\in\mathcal{E}^{\mathrm{ESW}}_g}\Delta u^{\mathrm{ESW}}_{kl,t} &\le u^{\mathrm{BB}}_{g,t - 1}M + 1,\label{eq:segmentstatus5}
    \end{align}
\end{subequations}
where $u^{\mathrm{BB}}_{g,t}$ is a binary variable indicating the energized status of bus block $g$ at time $t$, $u^{\mathrm{B}}_{b,t}$ denotes the energized status of bus $b$, $u^{\mathrm{E}}_{kl,t}$ represents the energized status of the line $(k,l)$, $M$ is a sufficiently large positive number, and $\mathcal{B}_g$ denotes the set of buses associated with bus block $g$.

Eq.~\eqref{eq:segmentstatus1} enforces monotonic energization, ensuring that once a bus block is energized, it remains energized throughout the entire horizon. Eqs.~\eqref{eq:segmentstatus2} and~\eqref{eq:segmentstatus3} ensure consistency by requiring that all buses and non-switchable lines within an energized bus block share the same status. Finally, Eqs.~\eqref{eq:segmentstatus4} and~\eqref{eq:segmentstatus5} enforce that at most one ESW may be closed to energize a de-energized bus block at any time step.

The GFMI provides the frequency reference for the MG it forms. From the beginning of the BS process, the frequency of the bus connected to the GFMI is propagated to the entire associated bus block according to:
\begin{equation}
    f_{b,t} = f_{g,t}, \forall b\in\left(\mathcal{B}^{\mathrm{GFMI}}_g\right),\forall g\in\mathcal{G},
\end{equation}
where $f_{b,t}$ and $f_{g,t}$ represent the frequency of bus $b$ and bus block $g$ at time $t$, respectively. The set $\mathcal{B}^{\mathrm{GFMI}}_g = \mathcal{B}^{\mathrm{GFMI}}\cap\mathcal{B}_g$ denotes the GFMI buses located within bus block $g$.

Additionally, once a bus block is energized, all buses at the terminals of switches within the same bus block must share the same frequency. This requirement is enforced by:
\begin{equation}
    f_{b,t} = f_{g,t}, \forall b\in\mathcal{B}^{\mathrm{SW}}_g, \forall g\in\mathcal{G},
\end{equation}
where $\mathcal{B}^{\mathrm{SW}}_g = \mathcal{B}^{\mathrm{SW}}\cap\mathcal{B}_g$ is the subset of switch terminal buses within bus block $g$.
\subsubsection{Energizing Switches (ESWs) Action}
For each ESW $(k,l)\in\mathcal{E}^{\mathrm{ESW}}$, its operational constraints are defined as follows:
\begin{subequations}\label{eq:eswaction}
    \begin{align}
    u^{\mathrm{ESW}}_{kl,t} &\le u^{\mathrm{B}}_{k,t - 1} + u^{\mathrm{B}}_{l,t - 1},\label{eq:eswaction1}\\
    \Delta u^{\mathrm{ESW}}_{kl,t} &\le 2 - u^{\mathrm{B}}_{k,t - 1} - u^{\mathrm{B}}_{l,t - 1},\label{eq:eswaction2}
    \end{align}
\end{subequations}

These constraints ensure that an ESW $(k,l)$ can only be used to energize a de-energized bus block if at least one of its terminal buses is already energized by a MG. However, it cannot be used to synchronize two already-operational MGs, as enforced by Eq.~\eqref{eq:eswaction2}.

Furthermore, when an ESW $(k,l)\in\mathcal{E}^{\mathrm{ESW}}$ is closed to pick up a de-energized bus block, the frequency of the MG associated with the energized bus (e.g., bus $k$) is transferred to the newly energized bus (e.g., bus $l$). This is captured by the following constraint:
\begin{equation}\label{eq:eswfrequency}
        f_{l,t} \hspace{-0.2em}-\hspace{-0.2em} \left(1 \hspace{-0.2em}-\hspace{-0.2em}u^{\mathrm{ESW}}_{kl,t}\right)M \le f_{k,t} \le f_{l,t} + \left(1\hspace{-0.2em} - \hspace{-0.2em}u^{\mathrm{ESW}}_{kl,t}\right)M,
\end{equation}

In addition, the constraint defined in Eq.~\eqref{eq:switchblockingfunction}, which governs the switch blocking function, is dynamically provided by the inrush current feasibility module to the controller. This feedback mechanism restricts the operation of ESWs to prevent unintended tripping of the MG recloser.
\subsubsection{Synchronizing Switches (SSWs) Action}
The SSW enables the coordination of two MGs during the BS, Its operation is subject to the following constraints:
\begin{subequations}\label{eq:sswaction}
    \begin{align}
    u^{\mathrm{SSW}}_{kl,t} &\ge u^{\mathrm{SSW}}_{kl,t - 1},\label{eq:sswaction1}\\
    2u^{\mathrm{SSW}}_{kl,t} &\le u^{\mathrm{B}}_{k,t - 1} + u^{\mathrm{B}}_{l,t - 1},\label{eq:sswaction2}\\
    f_{k,t,o} &\le f_{l,t,o} + \left(1 - u^{\mathrm{SSW}}_{kl,t}\right)M + \frac{\epsilon}{2},\label{eq:sswaction3}\\
    f_{k,t,o} &\ge f_{l,t,o} - \left(1 - u^{\mathrm{SSW}}_{kl,t}\right)M - \frac{\epsilon}{2},\label{eq:sswaction4}
    \end{align}
\end{subequations}
where $u^{\mathrm{SSW}}_{kl,t}$ is a binary variable indicating whether the SSW $(k,l)$ is closed at time $t$, and $\epsilon$ is a small positive number to define an acceptable synchronization tolerance.

Eq.~\eqref{eq:sswaction1} shows that the action of SSW is irreversible. Eq.~\eqref{eq:sswaction2} requires that both terminal buses of the SSW be energized prior to synchronization. Unlike ESWs, which transfer frequency from an energized segment to a de-energized one shown in Eq.~\eqref{eq:eswfrequency}, the SSW can only be closed if the frequencies at both terminals are synchronized, as enforced by Eqs.~\eqref{eq:sswaction3} and~\eqref{eq:sswaction4}.
\subsubsection{Radiality of the DS}
To ensure safe and stable operation during the BS, the radial topology of DS must be maintained. This requirement is dynamically enforced through the following constraints:
\begin{subequations}\label{eq:radiality}
    \begin{align}
    \sum_{\left(k,l\right)\in\mathcal{E}}u^{\mathrm{E}}_{kl,t} &= \sum_{b\in\mathcal{B}}u^{\mathrm{B}}_{b,t} - R_{t},\label{eq:radiality1}\\
    R_{t} &= |\mathcal{B}^{\mathrm{GFMI}}| + \sum_{b\in\mathcal{B}^{\mathrm{TG}}}u^{\mathrm{B}}_{i,t} \hspace{-0.2em}-\hspace{-1em} \sum_{(k,l)\in\mathcal{E}^\mathrm{SSW}}u^{\mathrm{SSW}}_{kl,t},\label{eq:radiality2}
    \end{align}
\end{subequations}
where $R_{t}$ denotes the dynamic number of root buses at time $t$, $\mathcal{B}^{\mathrm{TG}}$ represents the set of TG buses, and $\mathcal{E}^\mathrm{SSW}$ is the set of SSWs. Eq.~\eqref{eq:radiality1} ensures radiality by balancing the number of energized edges and buses, while Eq.~\eqref{eq:radiality2} defines $R_{t}$ based on the active roots at each time step of the BS.
\subsection{Formulation of System Operation}
\subsubsection{TG Outage and Recovery}
The TG is initially unavailable during the BS process until the faulted transmission line is repaired. The each TG-connected bus $b\in\mathcal{B}^{\mathrm{TG}}$, its output and operational status at each time step $t\in\mathcal{T}_c$ are governed by a binary indicator $u{\mathrm{B}}_{b,t}$, with the following constraints:
\begin{subequations}
    \begin{align}
    \boldsymbol{v}_{b,t} &= u^{\mathrm{B}}_{b,t}\boldsymbol{1}_{|\Phi|},\\ 
    f_{b, t} &= 60u^{\mathrm{B}}_{b,t},\\
    \left(\tfrac{1}{3}S^{\mathrm{TG}}_{b, \max}\right)^2 &\ge \max_{n\in\Phi}\left\{\left(p^{\mathrm{TG}}_{b,n,t}\right)^2 +\left(q^{\mathrm{TG}}_{b,n,t})^2\right)\right\},
    \end{align}
where variables $p^{\mathrm{TG}}_{b,n,t}$ and $q^{\mathrm{TG}}_{b,n,t}$ denote the active and reactive power output of the TG at phase $n$, respectively. $S^{\mathrm{TG}}_{b, max}$ is the maximum rated apparent power of the TG at bus $b$.
\end{subequations}
\subsubsection{Model of grid-forming inverters (GFMIs)}
The output of the GFMI-based baterry energy storage system (BESS) at bus $b\in\mathcal{B}^{\mathrm{GFMI}}$ is constrained by the rated apparent power and capacity at each time step $t\in\mathcal{T}_c$, which is illustrated as follows,
\begin{subequations}\label{eq:BESScontrol}
\begin{align}
    SoC_{b,t} &= SoC_{b,t-1} - \boldsymbol{1}_{|\Phi_i|}^{T}\boldsymbol{p}^{\mathrm{BESS}}_{b,t}/E^{\mathrm{BESS}}_{b,nom},\\
    \lfloor SoC \rfloor &\le SoC_{b,t} \le \lceil SoC \rceil,\label{eq:SoCconstraint}\\
    \left(\tfrac{1}{3}S^{\mathrm{BESS}}_{b, nom}\right)^2&\ge\max_{n\in\Phi}\left\{\left(p^{\mathrm{BESS}}_{b,n,t}\right)^2 +\left(q^{\mathrm{BESS}}_{b,n,t})^2\right)\right\},
\end{align}
\end{subequations}
where $SoC_{b,t}$ is the state of charge (SoC) of the BESS at bus $b$, variables $p^{\mathrm{BESS}}_{b,n,t}$ and $q^{\mathrm{BESS}}_{b,n,t}$ denote the active and reactive power output of the GFMI-based BESS at phase $n$, separately.

For the virtual synchronous generator (VSG)-controlled, GFMI-based BESS located at bus $b\in\mathcal{B}^{\mathrm{GFMI}}$, the frequency response to active power disturbances at time $t\in\mathcal{T}_c$ considering synchronization signals from SSWs, is modeled using a quasi-steady state (QSS) approximation. This is expressed as:
\begin{equation}
    f_{b,t} = f^{\mathrm{QSS}}_{b,t} + \Delta u^{\mathrm{SSW}}_{kl,t}\Delta f^{syn}_{b,t}, \forall (k,l)\in \mathcal{E}^{\mathrm{SSW}}
\end{equation}
where $f^{\mathrm{QSS}}_{b,t}$ denotes the QSS frequency of GFMI $b$, and $\Delta f^{syn}_{b,t}$ represents the synchronization-induced frequency adjustment. The detailed computation of $f^{\mathrm{QSS}}_{b,t}$, along with additional dynamic frequency indices, such as the maximum rate of change (RoC) of frequency $f_{b,t,\max}^{\mathrm{RoC}}$ and the frequency nadir $f^{nadir}_{b,t}$, can be found in our prior work~\cite{Maharjan2025}.

To ensure frequency security during both the transient and steady-state phases of the BS process, the following constraints are imposed on the GFMI frequency and its associated indices:
\begin{subequations}\label{eq:frequencysecurity}
    \begin{align}
    \lfloor f_k \rfloor &\le f_{k,t,o} \le \lceil f_k \rceil\label{eq:frequencysecurityfrequency},\\
        \lfloor f^{\mathrm{QSS}} \rfloor &\le f^{\mathrm{QSS}}_{k,t,o} \le \lceil f^{\mathrm{QSS}} \rceil\label{eq:frequencysecurityQSS},\\
    \lfloor f^{\mathrm{RoC}}_{\max} \rfloor &\le f^{\mathrm{RoC}}_{k,t,o,\max} \le \lceil f^{\mathrm{RoC}}_{\max} \rceil\label{eq:frequencysecurityRoC},\\
    \lfloor f^{\mathrm{nadir}} \rfloor &\le f^{\mathrm{nadir}}_{k,t,o} \le \lceil f^{\mathrm{nadir}}\rceil\label{eq:frequencysecuritynadir}.
    \end{align}
\end{subequations}

Moreover, the output voltage of the GFMI is regulated by the voltage reduction function whenever a melted fused is detected within the associated MG by the inrush current feasibility module, as specified in Eqs.~\eqref{eq:gfmivoltage_r}-\eqref{eq:gfmivoltage_l}.
\subsubsection{Model of photovoltaics (PV)}
The active power output of a behind-the-meter (BTM) PV system is subject to environmental variability, while its reactive power output is typically maintained at a fixed ratio relative to the active component. Let $\eta_{t}$ denote the normalized PV output rate at time $t$, and let $\mathcal{B}^\mathrm{PV}$ represent the set of buses equipped with PV systems. For each bus $b\in\mathcal{B}^\mathrm{PV}$, the output of the PV system is constrained by:
\begin{subequations}\label{eq:PVcontrol}
\begin{align}
    \boldsymbol{p}^{\mathrm{PV}}_{b,t} &= \tfrac{1}{3}S^{\mathrm{PV}}_{b,nom}\eta_{t}\boldsymbol{1}_{|\Phi_b|},\\
    \boldsymbol{q}^{\mathrm{PV}}_{b,t} &= 0.352\boldsymbol{p}^{\mathrm{PV}}_{b,t},
\end{align}
\end{subequations}
where $S^{\mathrm{PV}}_{b,nom}$ is the nominal apparent power rating of the PV system at bus $b$.
\subsubsection{ZIP Loads Considering Cold Load Pick-up Effect}
\paragraph{Modeling of ZIP loads}
ZIP models, as established in~\cite{Ma2022}, are widely adopted by power utilities to characterize load behavior. For each bus $b\in\mathcal{B}^{\mathcal{X}}$, we redefine the active and reactive ZIP load, denoted by $\boldsymbol{p}_{b,t}$ and $\boldsymbol{q}_{b,t}$, in terms of the squared voltage magnitude to align with the branch flow model of the DS. The linearized forms of these models are expressed as follows:
\begin{subequations}\label{eq:ZIP_load}
\begin{align}
  \boldsymbol{p}_{b,t}^{\mathrm{ZIP}} &\hspace{-0.4em}= \hspace{-0.2em}\boldsymbol{p}_{b,t}^{\mathrm{L}}\hspace{-0.2em}\circ\hspace{-0.4em}\left[\hspace{-0.2em}\left(\boldsymbol{k}^p_{\mathrm{Z},b}\hspace{-0.2em} + \hspace{-0.2em}\boldsymbol{k}^p_{\mathrm{I},b}\hspace{-0.2em}\circ\hspace{-0.2em}\frac{1}{2\sqrt{\boldsymbol{v}_{b,m}}}\hspace{-0.2em}\right)\hspace{-0.4em}\circ\hspace{-0.2em}\boldsymbol{v}_{b,t}\hspace{-0.2em}+\hspace{-0.2em}\boldsymbol{k}^p_{\mathrm{I},b}\hspace{-0.2em}\circ\hspace{-0.2em}\frac{\sqrt{\boldsymbol{v}_{b,m}}}{2}\hspace{-0.2em}+\hspace{-0.2em} \boldsymbol{k}^p_{\mathrm{P},b}\hspace{-0.2em}\right]\hspace{-0.2em},\\
  \boldsymbol{q}_{b,t}^{\mathrm{ZIP}} &\hspace{-0.4em}= \hspace{-0.2em}\boldsymbol{q}^{\mathrm{L}}_{b,t}\hspace{-0.2em}\circ\hspace{-0.4em}\left[\hspace{-0.2em}\left(\boldsymbol{k}^q_{\mathrm{Z},b} \hspace{-0.2em}+ \hspace{-0.2em}\boldsymbol{k}^q_{\mathrm{I},b}\hspace{-0.2em}\circ\hspace{-0.2em}\frac{1}{2\sqrt{\boldsymbol{v}_{b,m}}}\hspace{-0.2em}\right)\hspace{-0.4em}\circ\hspace{-0.2em}\boldsymbol{v}_{b,t}\hspace{-0.2em}+ \hspace{-0.2em}\boldsymbol{k}^q_{\mathrm{I},b}\hspace{-0.2em}\circ\hspace{-0.2em} \frac{\sqrt{\boldsymbol{v}_{b,m}}}{2}\hspace{-0.2em}+ \hspace{-0.2em}\boldsymbol{k}^q_{\mathrm{P},b}\hspace{-0.2em}\right]\hspace{-0.2em},
\end{align}
\end{subequations}
where $\boldsymbol{p}_{b,t}^{\mathrm{L}}$ and $\boldsymbol{q}^{\mathrm{L}}_{b,t}$ are column vectors representing the nominal active and reactive load demands at bus $b$, respectively. The vectors $\boldsymbol{k}_{\mathrm{Z},b}$, $\boldsymbol{k}_{\mathrm{I},b}$, and $\boldsymbol{k}_{\mathrm{P},b}$ contain the ZIP model coefficients corresponding to the constant impedance, constant current, and constant power components at bus $b$.

Please refer to Appendix~\ref{app_lin_ZIP} for the derivation of Eq.~\eqref{eq:ZIP_load}.
\paragraph{Modeling of cold load pick-up (CLPU) effect}
During BS, the initial load demand may surge to several times the nominal level due to the simultaneous reactivation of thermal and motor loads. This phenomenon, known as the CLPU effect, poses a significant challenge during the system restoration.

To capture the dynamic behavior of CLPU, a staircase-based model is introduced. The actual active and reactive power demands of CL at each bus $b\in\mathcal{B}^{\mathrm{CL}}_g$ within bus block $g\in\mathcal{G}$, are formulated as:
\begin{subequations}\label{eq:clpu_cl}
\begin{align}
    \boldsymbol{p}_{b,t}^{\mathrm{CL}} = &\;\boldsymbol{p}^{\mathrm{ZIP}}_{b,t}\left[\textstyle\sum_{o=1}^{3}\left(\beta_o \Delta u^{\mathrm{BB}}_{g,t-(o-1)}\right) + u^{\mathrm{BB}}_{g,t}\right],\\
    \boldsymbol{q}_{b,t}^{\mathrm{CL}} = &\;\boldsymbol{p}_{b,t}^{\mathrm{CL}}\tan\left(\varphi^{\mathrm{CL}}_b\right),
\end{align}
\end{subequations}
where $\beta_1$, $\beta_2$, and $\beta_3$ are CLPU amplification coefficients, and $\varphi^{\mathrm{CL}}_b$ is the power factor angle of the CL at bus $b$.

Similarly, the restored active and reactive power of NL at each bus $b\in\mathcal{B}^{\mathrm{NL}}_g$ and bus block $g\in\mathcal{G}$, are modeled as:
\begin{subequations}\label{eq:clpu_nl}
\begin{align}
    \boldsymbol{p}_{b,t}^{\mathrm{NL}} = &\;\boldsymbol{p}^{\mathrm{ZIP}}_{b,t}\left[\textstyle\sum_{o=1}^{3}\left(\beta_o \Delta w^{\mathrm{B}}_{b,t-(o-1)}\right) + w^{\mathrm{B}}_{b,t}\right],\\
    \boldsymbol{q}_{b,t}^{\mathrm{NL}} = &\;\boldsymbol{p}_{b,t}^{\mathrm{NL}}\tan\left(\varphi^{\mathrm{NL}}_b\right),\\
    w^{\mathrm{B}}_{b,t} \le &\; u^{\mathrm{BB}}_{g,t},\\
    w^{\mathrm{B}}_{b,t - 1}\le &\;w^{\mathrm{B}}_{b,t},
\end{align}
\end{subequations}
where $w^{\mathrm{B}}_{b,t}$ is a binary variable indicating the energized status of the NL at bus $b$, which is dominated by the bus block status.
\subsubsection{Unbalanced Linear Power Flow}
Define the set of child buses of bus $b$ as $\mathcal{B}_b^{ch}$, the nodal power balance constraints for each bus $b\in\mathcal{B}$ at time $t\in\mathcal{T}_c$ are given by:
\begin{subequations}\label{eq:power_flow_unbalanced}
   \begin{align}   
    \boldsymbol{p}_{b,t} &= \textstyle\sum_{k\in \mathcal{B}_b^{ch}}\left(\boldsymbol{\Lambda}_{|\Phi_{ab}|\times|\Phi_{bk}|}\boldsymbol{p}_{bk,t}\right) - \boldsymbol{p}_{ab,t}\\
    \boldsymbol{q}_{b,t} &= \textstyle\sum_{k\in \mathcal{B}_b^{ch}}\left(\boldsymbol{\Lambda}_{|\Phi_{ab}|\times|\Phi_{bk}|}\boldsymbol{q}_{bk,t}\right) - \boldsymbol{q}_{ab,t},\\
    \boldsymbol{p}_{b,t} &= \boldsymbol{p}^{\mathrm{TG}}_{b,t} + \boldsymbol{p}^{\mathrm{BESS}}_{b,t} + \boldsymbol{p}^{\mathrm{PV}}_{b,t} - \boldsymbol{p}^{\mathrm{CL}}_{b,t} - \boldsymbol{p}^{\mathrm{NL}}_{b,t}\\
    \boldsymbol{q}_{b,t} &= \boldsymbol{q}^{\mathrm{TG}}_{b,t} + \boldsymbol{q}^{\mathrm{BESS}}_{b,t} + \boldsymbol{q}^{\mathrm{PV}}_{b,t} - \boldsymbol{q}^{\mathrm{CL}}_{b,t} - \boldsymbol{q}^{\mathrm{NL}}_{b,t}
\end{align}
\end{subequations}
where $\boldsymbol{p}_{b,t}$ and $\boldsymbol{q}_{b,t}$ are column vectors representing the net active and reactive power injection at bus $b$, respectively. Bus $a$ denotes the parent of bus $b$, and $\boldsymbol{\Lambda}_{|\Phi_{ab}|\times|\Phi_{bk}|}$ is a binary mapping matrix of size $|\Phi_{ab}|\times|\Phi_{bk}|$ with all one rows corresponding to the phases in $\Phi_k$.

The voltage relationship between the sending and receiving ends of each edge $(k,l)\in\mathcal{E}$ at time $t\in\mathcal{T}_c$ is enforced through the following constraints:
\begin{subequations}\label{eq:line_flow_unbalanced}
\begin{align}
    \boldsymbol{v}_{l,t}\le &\left[\boldsymbol{v}_{k,t}^{\Phi_{kl}} - 
    2\left(\boldsymbol{r}_{kl}\boldsymbol{p}_{kl,t} + \boldsymbol{x}_{kl}\boldsymbol{q}_{kl,t}\right)\right.\notag\\
    &\left.+ \left(1-u^{\mathrm{E}}_{kl,t}\right)\boldsymbol{M}^{\Phi_{kl}}\right],\label{eq:line_flow_r}\\
    \boldsymbol{v}_{l,t} \ge &\left[\boldsymbol{v}_{k,t}^{\Phi_{kl}} - 
    2\left(\boldsymbol{r}_{kl}\boldsymbol{p}_{kl,t} + \boldsymbol{x}_{kl}\boldsymbol{q}_{kl,t}\right)\right.\notag\\
    &\left.- \left(1-u^{\mathrm{E}}_{kl,t}\right)\boldsymbol{M}^{\Phi_{kl}}\right],\label{eq:line_flow_l}
\end{align}
\end{subequations}
where '$\cdot^{\Phi_{kl}}$' denotes the extraction of phase components relevant to edge $(k,l)$, and $\boldsymbol{r}_{kl}$, $\boldsymbol{x}_{kl}$ are matrices associated with branch resistance and reactance, whose calculations are described in~\cite{Cheng2022}.

To ensure the operational security of the restored bus blocks, the following constraints are imposed for each edge $(k,l)\in\mathcal{E}$ and $b\in\mathcal{B}$:
\begin{subequations}\label{eq:powerflowsecurity}
    \begin{align}
    -u^{\mathrm{E}}_{kl,t}\boldsymbol{M}^{\Phi_{kl}} &\le \boldsymbol{p}_{kl,t} \le u^{\mathrm{E}}_{kl,t}\boldsymbol{M}^{\Phi_{kl}},\\
    -u^{\mathrm{E}}_{kl,t}\boldsymbol{M}^{\Phi_{kl}} &\le \boldsymbol{q}_{kl,t} \le u^{\mathrm{E}}_{kl,t}\boldsymbol{M}^{\Phi_{kl}},\\  
    u_{b,t}^{\mathrm{B}}\lfloor \boldsymbol{v}_{b} \rfloor &\le \boldsymbol{v}_{b,t} \le u_{b,t}^{\mathrm{B}}\lceil \boldsymbol{v}_{b} \rceil\label{eq:powerflowsecurity3},
    \end{align}
\end{subequations}
where the nodal voltage lower bound may be adaptively reduced if the voltage reduction mechanism is triggered by the inrush current feasibility module.
\subsection{Solution Structure of The MPBS Framework}
To implement the proposed restoration strategy in real time, the MPBS framework is developed, as outlined in Algorithm~\ref{Alg_MPBS}. This framework integrates short-term forecasting, optimization, and inrush current feasibility checks into a closed-loop control architecture, ensuring that each restoration action is both optimal and feasible under dynamic system conditions.
\begin{algorithm}[t]
\color{black}
\caption{The Proposed MPBS Framework} \label{Alg_MPBS}
\LinesNumbered
\SetAlgoLined
\DontPrintSemicolon
\textbf{Initialization:} Set $t \leftarrow 1$; initialize system status (energized buses, switches, frequencies, indicators).\;
\textbf{Forecasting:} Predict DER output and TG availability over controller horizon $\mathcal{T}_c$.\;
\textbf{Optimization:} Solve the MPBS problem (objective: Eq.~\eqref{eq:objecivefunction}; constraints: Eqs.~\eqref{eq:segmentstatus}–\eqref{eq:powerflowsecurity}).\;
\textbf{Inrush Current Feasibility:} If inrush check fails: activate voltage reduction (Eqs.~\eqref{eq:gfmivoltage_r}–\eqref{eq:gfmivoltage_l}), block high-risk switches (Eq.~\eqref{eq:switchblockingfunction}), and re-solve.\;
\textbf{Execution:} Implement first-step control actions.\;
\textbf{Update:} Refresh system state (frequency, voltage, load modeling).\;
\textbf{Advance:} $t \leftarrow t+1$; repeat until restoration completes or $t > \max(\mathcal{T})$.\;
\textbf{Output:} Return real-time feasible and optimal restoration sequence.
\end{algorithm}
\section{Results}\label{sec:v}
This section begins by introducing the test system along with its detailed parameters. Next, the proposed inrush current model for switching a no-load DT is validated against EMT simulations conducted in PowerFactory, where various closing phase angles relative to phase A are applied. The subsequent part presents the BS process of the system, with a focus on the impact of inrush current on protection device behavior. Finally, the effectiveness of energy conservation strategies during the BS process is demonstrated and analyzed.
\subsection{Simulation Setup}
\begin{figure}[htbp]
    \centering
    \includegraphics[width=\linewidth]{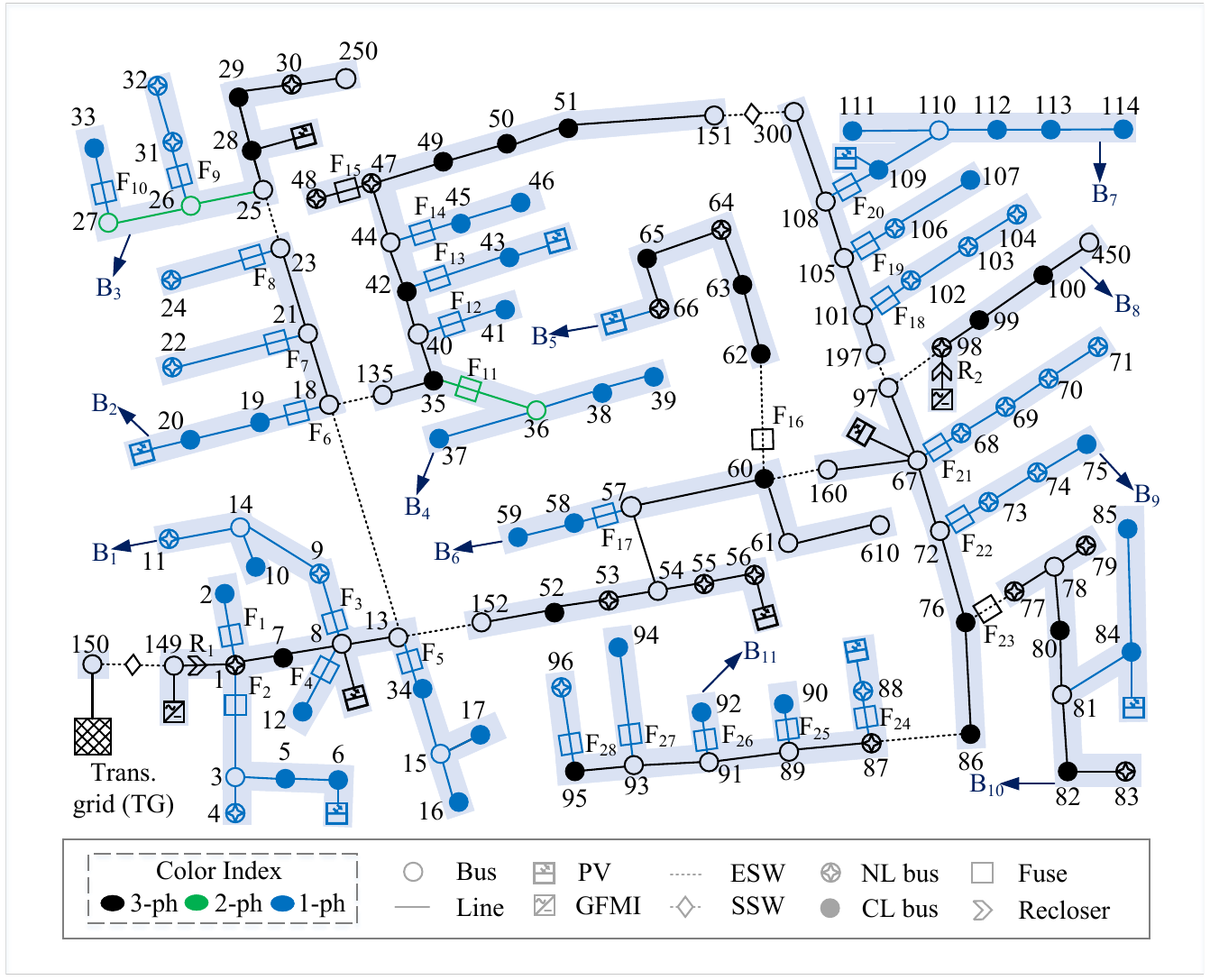}
    \vspace{-2em}
    \caption{Modified IEEE 123-node feeder with protection devices.}
    \label{fig:testedfeeder}
\end{figure}

The test system is adapted from the three-phase unbalanced IEEE 123-node feeder, incorporating protection devices such as fuses and reclosers as described in~\cite{ButlerPurry2009}. The system layout is illustrated in Fig.~\ref{fig:testedfeeder}, where colors denote the number of phases and symbols indicate various devices. Among the buses, 88 are connected to loads, with 52 categorized as CLs. The system's peak active power demand is approximately 3.49 MW, assuming a uniform power factor of $\varphi=0.484$ across the DS. BTM PV systems are installed with a total capacity of $965$ kW, accounting for $28\%$ of the system's peak demand. Each load is connected through a single-phase DT, whose capacity is selected based on the corresponding maximum load demand. The detailed parameters about DT in shown in Appendix~\ref{DTParameters}.

Two GFMI-based BESSs are located at buses $149$ and $98$. Their rated apparent powers are $2.294$ MW and $2.222$ MW, respectively, with corresponding energy capacities of $3.942$ MWh and $3.587$ MWh. The DS is partitioned into $11$ bus blocks, denoted as $\mathrm{B}_g,g\in\mathcal{G}$, using $10$ ESWs and $2$ SSWs. Additionally, a TG connection is located at bus $150$. The security limits for frequency-related metrics are shown in Appendix~\ref{frequencylimits}.
\subsection{Verification of Inrush Current Estimation}
The test DS with GFMI-based BESS is modeled in PowerFactory, incorporating the detailed saturation characteristics of DTs. EMT simulations are conducted by applying four different closing angles--$0^\circ$, $30^\circ$, $60^\circ$, and $90^\circ$ (all referenced to phase A)--to the high-voltage side of a randomly selected $100$ kVA DT. The simulation results are presented below.

Fig.~\ref{fig:EMTInrushCurrentviaPhaseAngles}~(a)-(d) illustrate the real-time three-phase current waveforms for the first two cycles following DT energization under the four closing angle scenarios. As predicted by Eq.~\eqref{eq:saturationindicator}, with $\lambda_0 = 0.8$ and $\lambda_s = 1.2$ for phase A, all four closing angles lead to saturation of the DT's phase A winding in the positive direction. Consistent with Eq.~\eqref{eq:singleinrushcurrent}, the peak inrush current magnitude decreases as the closing angle increases.
\begin{figure}[htbp]
    \centering
    \includegraphics[width=\linewidth]{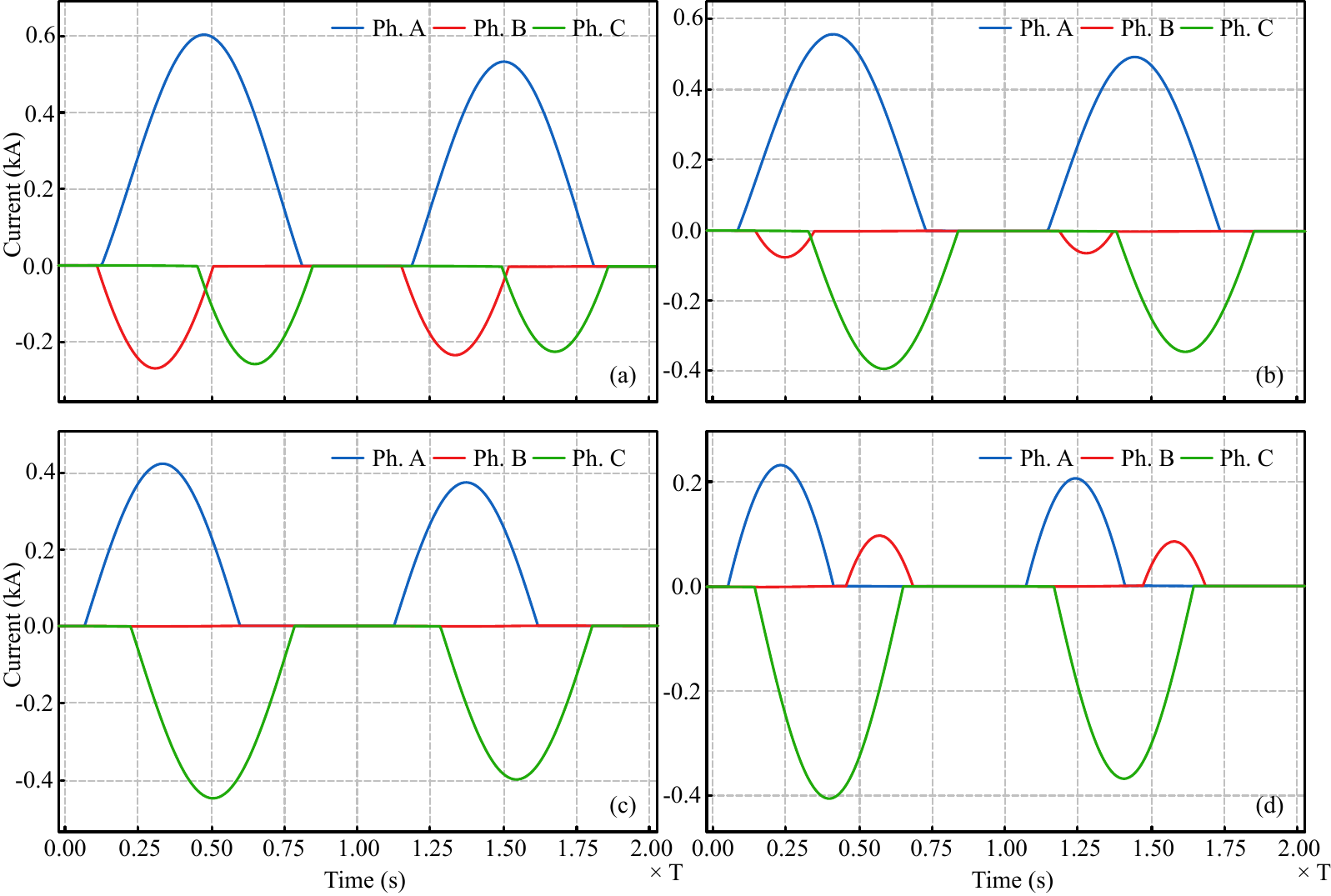}
    \vspace{-2em}
    \caption{Inrush Currents of EMT simulations at different closing phase angles.}
    \label{fig:EMTInrushCurrentviaPhaseAngles}
\end{figure}

To quantitatively assess the accuracy of the estimation, the simulation results in Fig.~\ref{fig:EMTInrushCurrentviaPhaseAngles} are summarized in Table~\ref{tab:VerificationofInrushCurrent}. The comparison between the estimated and measured peak inrush currents demonstrates high accuracy in estimation, with all cases exceeding $90\%$.
\begin{table}[htbp]
\centering
\caption{Verification of Inrush Current Estimation}\label{tab:VerificationofInrushCurrent}
\begin{tabular}{c c c c c}
\toprule
\multirow{2}{*}{\makecell{Closing\\angle}} & \multirow{2}{*}{Phase} & \multicolumn{2}{c}{$I_{\max}$ (kA)} & \multirow{2}{*}{Accuracy (\%)}\\
\cmidrule(lr){3-4}
 & & Est. & Mea. & \\
\midrule
$0^\circ$ & A & 0.6440 & 0.6035 & 93.29 \\
$30^\circ$ & A & 0.5901 & 0.5586 & 94.36 \\
$60^\circ$ & A & 0.4428 & 0.4224 & 95.17 \\
$90^\circ$ & C & 0.3759 & 0.4055 & 92.70 \\
\bottomrule
\end{tabular}
\end{table}
\subsection{Optimal BS Process Considering The Impact of Inrush Current on The Protection Devices}
\begin{figure}[htbp]
    \centering
    \includegraphics[scale=0.95]{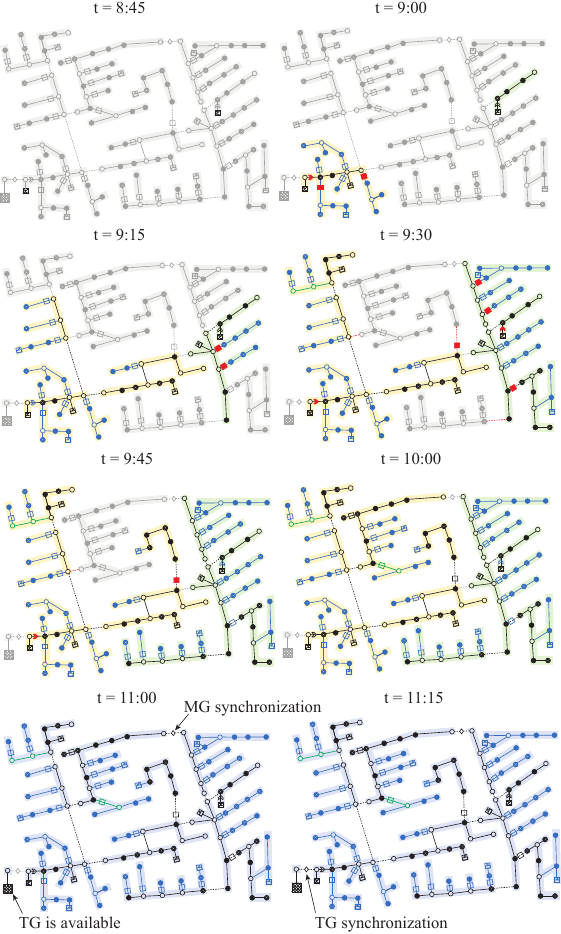}
    \vspace{-1em}
    \caption{Optimal BS cranking path considering the impact of inrush current on protection devices.}
    \label{fig:BSProcessConsideringInrushCurrent}
\end{figure}

The complete BS cranking path, incorporating the effects of inrush current on protection device, is illustrated in Fig.~\ref{fig:BSProcessConsideringInrushCurrent}. In this scenario, the TG remains unavailable until 11:00.

As shown in Fig.~\ref{fig:BSProcessConsideringInrushCurrent}, the BS process is initiated by energizing the two self-starting GFMI-based BESSs. At 9:00, the corresponding bus blocks $\mathrm{B}_1$ and $\mathrm{B}_8$, where the GFMIs are located, are energized and operate as two separate MGs. Within $\mathrm{B}_1$, two fuses, $\mathrm{F}_2$ and $\mathrm{F}_5$, are marked in red, indicating that the voltage reduction function was triggered to suppress excessive inrush current. As a result, the optimization problem was re-solved multiple times at this time step to prevent fuse misoperation.

At 9:15, both MGs expand by energizing additional bus blocks, $\mathrm{B}_2$, $\mathrm{B}_6$, and $\mathrm{B}_9$. Subsequently, at 9:30, further expansion brings $\mathrm{B}_3$, $\mathrm{B}_7$, and $\mathrm{B}_{10}$ online. However, both MG reclosers are flagged red, signaling that they experienced misoperation in earlier iterations of the optimization process at this step. To mitigate this, the algorithm blocks several high-risk ESWs, including $(18, 135)$, $(60, 62)$, and $(86, 87)$. 

By 10:00, all bus blocks are successfully energized. The two MGs remain independently operating until 11:00, when they are synchronized in preparation for reconnection with the now-available TG. At 11:15, the TG is synchronized with the combined MG, completing the restoration process. The system then maintains this steady-state configuration.

\begin{figure}[htbp]
    \centering
    \includegraphics[width=\linewidth]{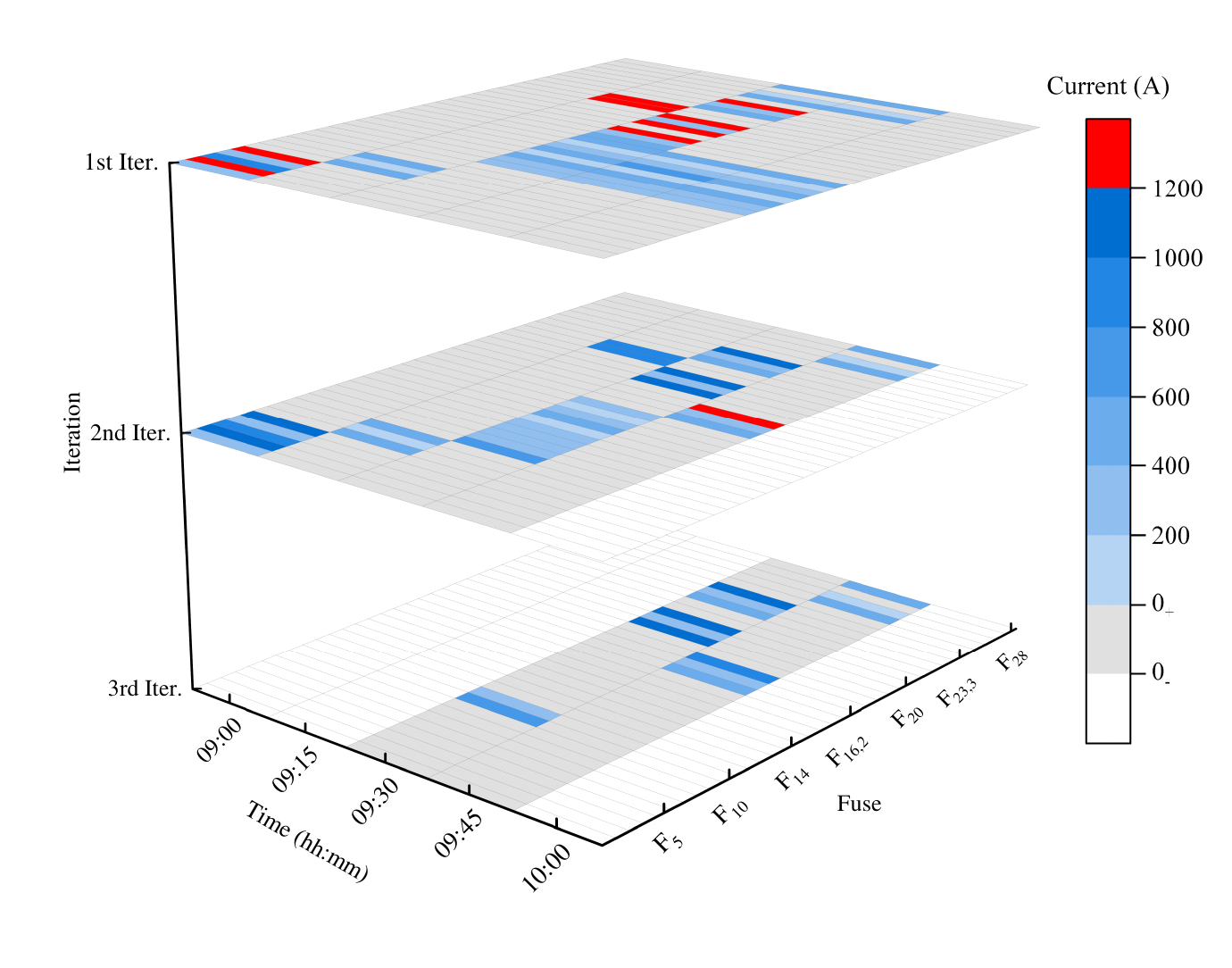}
    \vspace{-3em}
    \caption{Inrush currents through fuses during the BS process.}
    \label{fig:InrushCurrentThroughFuses}
\end{figure}

Detailed information on the inrush current flowing through each fuse during every iteration of the BS is presented in Fig.~\ref{fig:InrushCurrentThroughFuses}.
\vspace{-0em}

As shown in Fig.~\ref{fig:InrushCurrentThroughFuses}, each time step can have up to three optimization iterations, as observed at 9:30 and 9:45. Blank (white) areas indicate that no such iteration occurred at those time steps. Within a given iteration, gray cells signify that the corresponding fuse experienced no inrush current, typically because its associated bus block was not energized.

\begin{figure}[htbp]
    \centering
    \includegraphics[width=\linewidth]{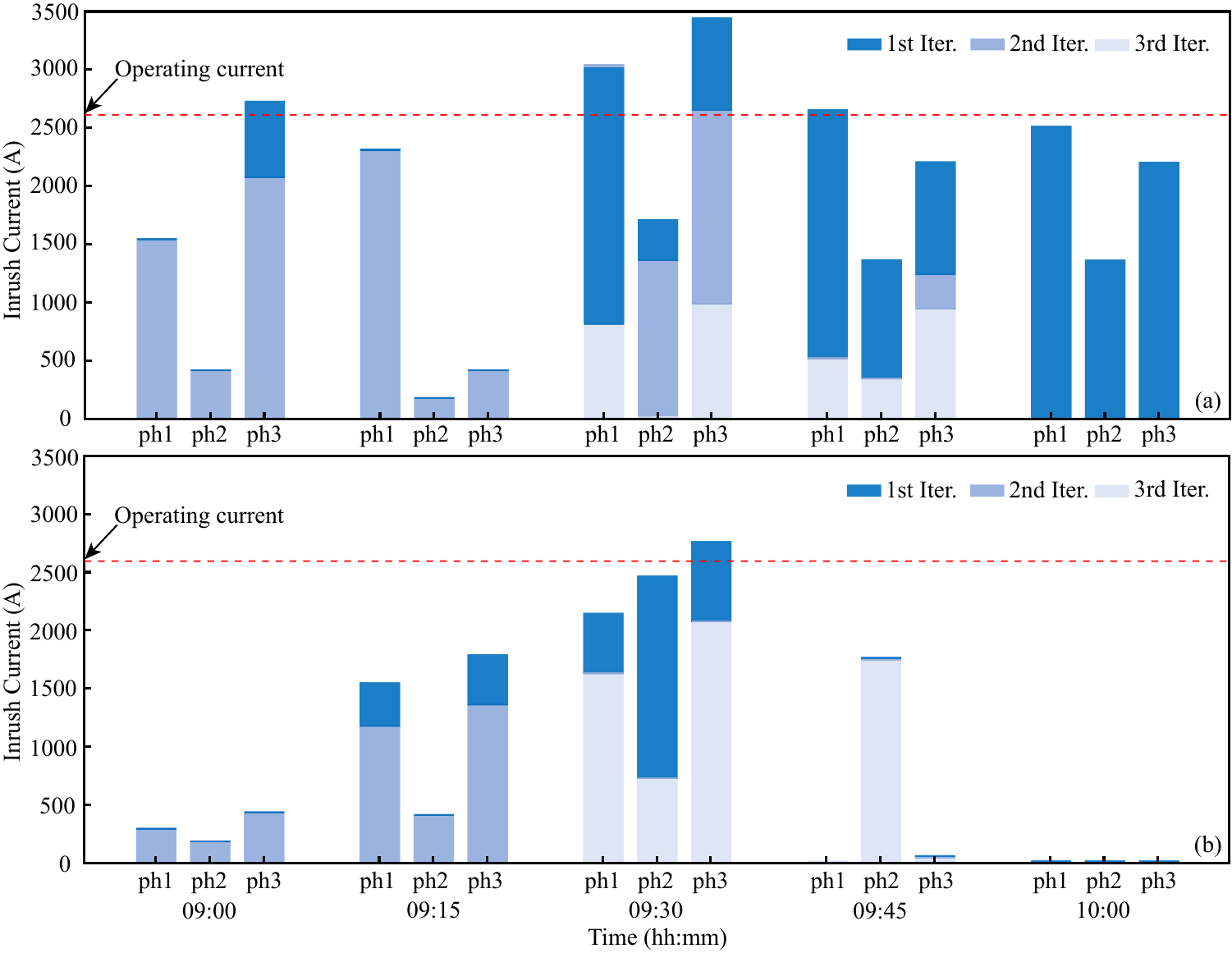}
    \vspace{-2em}
    \caption{Inrush currents through reclosers during the BS process}
    \label{fig:InrushCurrentThroughRecloser}
\end{figure}

The inrush current magnitude is color-coded: deeper shades of blue represent higher current levels, while red indicates that the fuse melted during that specific iteration. For instance, at 9:00 during the first iteration, fuse $\mathrm{F}_2$ experienced an inrush current of 1362.057 A, exceeding its minimum melting threshold of 1200 A (for a 2-cycle duration), resulting in fuse failure.
Similarly, the inrush current flowing through the reclosers during the BS process are shown in Fig.~\ref{fig:InrushCurrentThroughRecloser}.

As illustrated in Fig.~\ref{fig:InrushCurrentThroughRecloser}, both reclosers are triggered to misoperate during the first iteration at time step 9:30. Specifically, the inrush currents through phase 1 and phase 3 of recloser $\mathrm{R}_1$ reach 3013.166 A and 3440.931 A, respectively, while phase 3 of recloser $\mathrm{R}_2$ experiences 2077.432 A. These values exceed the reclosers’ minimum pickup current of 2600 A (for a 2-cycle duration), causing unintended tripping. To address this, the switch blocking function is activated through the inrush current feasibility module. After several iterations, the reclosers are successfully prevented from misoperating.
\subsection{Energy Conservation during The BS Process}
The effectiveness of the emergency operation-inspired voltage reduction strategy in conserving energy during the BS process is evaluated as follows.

\begin{figure}[htbp]
    \centering
    \includegraphics[width=\linewidth]{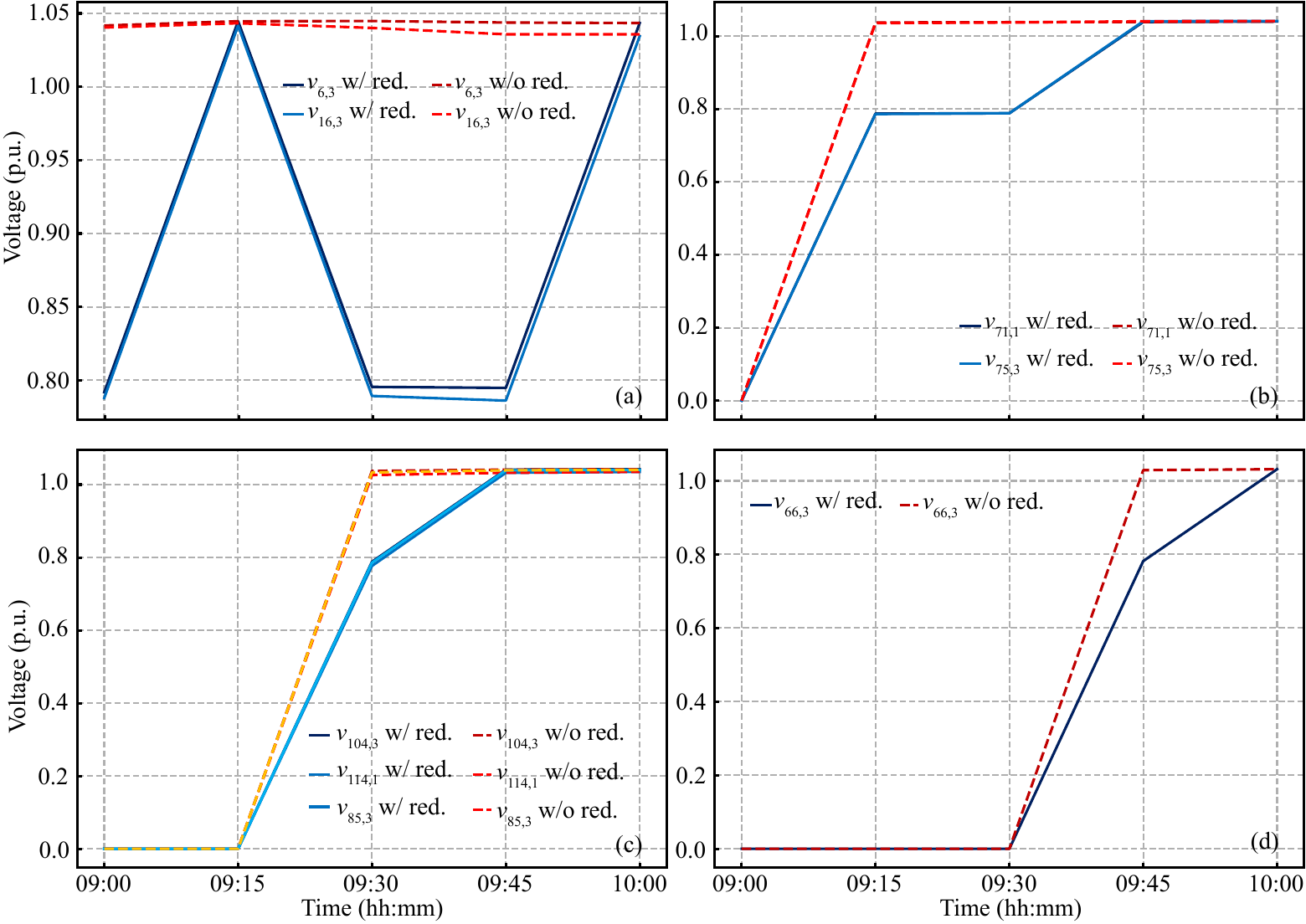}
    \vspace{-2em}
    \caption{Voltage with or without reduction during the BS process.}
    \label{fig:VoltageWorWOVR}
\end{figure}

Fig.~\ref{fig:VoltageWorWOVR} shows the voltage profiles at selected buses located along laterals where fuses are at risk of melting, under two scenarios: with and without the application of voltage reduction.

As depicted in Fig.~\ref{fig:VoltageWorWOVR}, nodal voltages at certain locations, such as phase 3 of bus 6 and bus 16, are deliberately reduced to 0.8 per unit to mitigate excessive inrush currents. This action is enabled by the voltage reduction function defined in Eq.~\eqref{eq:gfmivoltage}.

\begin{figure}[htbp]
    \centering
    \includegraphics[width=\linewidth]{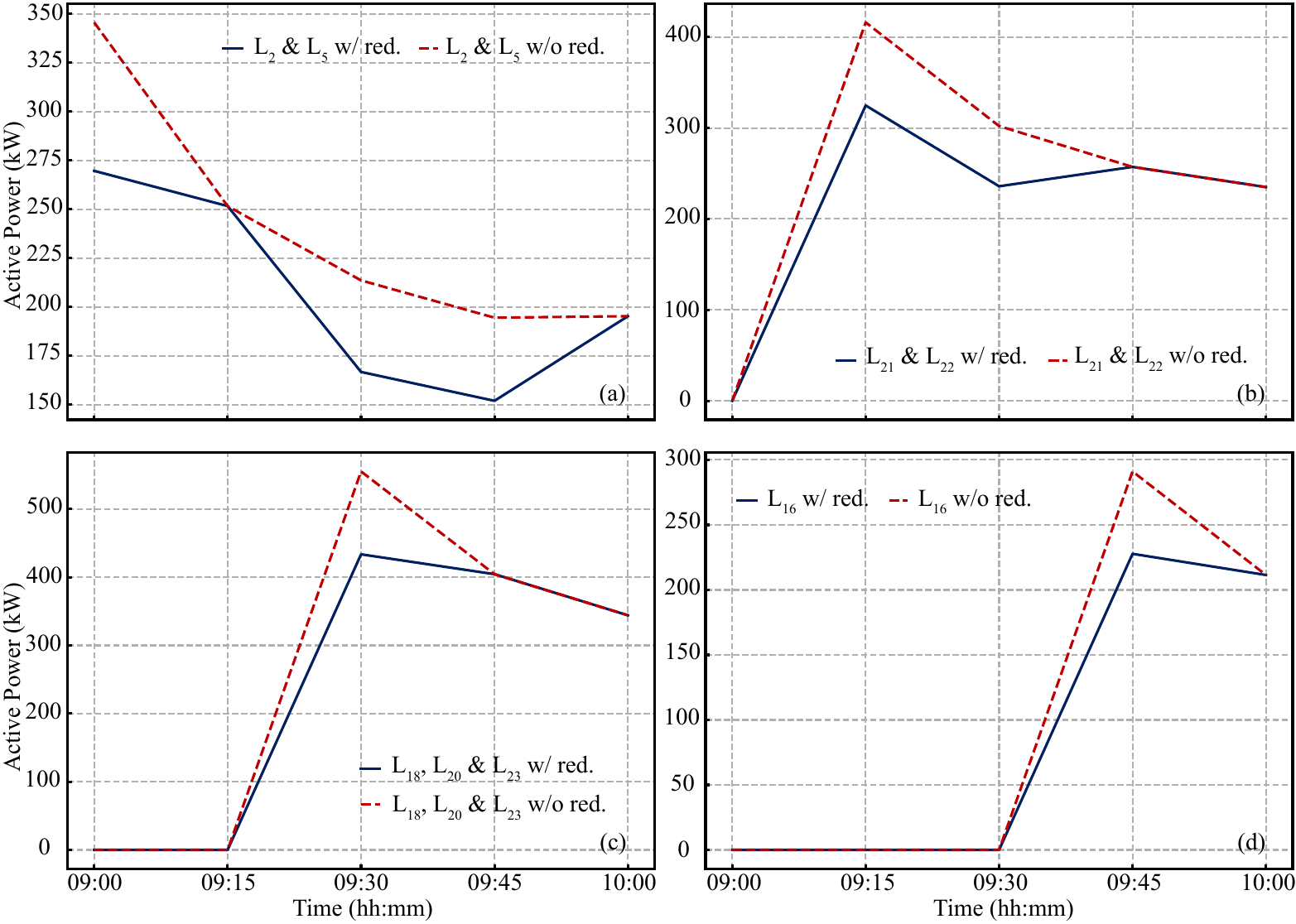}
    \vspace{-2em}
    \caption{Active power demands with or without voltage reduction during the BS process.}
    \label{fig:ActivePowerDemandWorWOVR}
\end{figure}

The impact of this voltage reduction on load demand is shown in Fig.~\ref{fig:ActivePowerDemandWorWOVR}, where the modeled ZIP loads also account for the CLPU effect. The reduced voltage leads to lower active power demands in the affected areas compared to the case without voltage reduction.

As shown, the application of voltage reduction results in noticeable power savings, particularly within the laterals impacted by voltage reduction. The cumulative energy consumption of the two GFMI-based BESSs located at buses 149 and 98 from 08:45 to 10:00 is summarized in Table~\ref{tab:BESSEnergyConsumption}. The energy savings attributed to voltage reduction during this period are approximately 76 kWh and 114 kWh for BESS 149 and BESS 98, respectively.
\begin{table}[htbp]
\centering
\caption{Energy Consumption of BESSs W/ or W/O Voltage Reduction}\label{tab:BESSEnergyConsumption}
\begin{tabular}{c c c c c}
\toprule
\multirow{2}{*}{\makecell{Time\\(hh:mm)}} & \multicolumn{2}{c}{$E_{\mathrm{149}}$ (MWh)} & \multicolumn{2}{c}{$E_{\mathrm{98}}$ (MWh)}\\
\cmidrule(lr){2-3} \cmidrule(lr){4-5}
 & w/ red. & w/o red. & w/ red. & w/o red. \\
\midrule
08:45 & 3.942 & 3.942 & 3.587 & 3.587\\
09:00 & 3.805 & 3.786 & 3.535 & 3.535\\
09:15 & 3.584 & 3.565 & 3.339 & 3.295\\
09:30 & 3.414 & 3.372 & 3.014 & 2.901\\
09:45 & 3.157 & 3.081 & 2.619 & 2.506\\
10:00 & 2.739 & 2.663 & 2.282 & 2.168\\
\bottomrule
\end{tabular}
\end{table}
\section{Conclusion}\label{sec:vi}
This paper presents a model predictive black start (MPBS) framework integrated with an inrush current feasibility module to enable real-time feasible and optimal DER-led restoration sequencing in the DS. Short-term forecasts of DER output and TG availability are generated by the DS plant and used by the controller to compute a rolling-horizon cranking path. At each time step, the inrush current feasibility module evaluates the estimated inrush current, where the emergency operation-inspired voltage reduction strategy and switch blocking mechanism will be activated if excessive inrush currents are detected. The key findings are as follows. First, the proposed analytical model accurately estimates inrush current during no-load DT energization, achieving over $90\%$ accuracy compared to EMT simulations in PowerFactory. Second, the MPBS framework successfully avoids protection device misoperations while optimizing the restoration sequence under dynamic conditions. Moreover, the integration of voltage-sensitive ZIP loads with voltage reduction enables load relief and energy savings during the BS. Future work will investigate the uncertainties in the real-time forecasting to further enhance the proposed framework.
\appendix
\subsection{Linearization of ZIP Loads} 
\label{app_lin_ZIP}
\begin{subequations}\label{eq:ZIP_load_model}
\begin{align}
  p^{\mathrm{ZIP}} &=  p^{\mathrm{L}}\left(k^p_\mathrm{Z} v +k^p_\mathrm{I} \sqrt{v}+ k^p_\mathrm{P}\right)\\
  q^{\mathrm{ZIP}} &=  q^{\mathrm{L}}\left(k^q_\mathrm{Z} v +k^q_\mathrm{I} \sqrt{v}+ k^q_\mathrm{P}\right)
\end{align}
\end{subequations}
Eq.~\eqref{eq:ZIP_load_model} is non-linear due to the square-root of $v$, which can be linearized in the vicinity of the present voltage measurement in p.u. ($v_m$) as:
\begin{equation}
    \sqrt{v} \approx \frac{v}{2\sqrt{v_m}}+\frac{\sqrt{v_m}}{2}
\end{equation}
\subsection{DT Parameters}\label{DTParameters}
At the onset of the BS, the initial residual flux values in per unit for phases A, B, and C are set to 0.8, –0.4, and –0.4, respectively. The saturated inductance of each DT is assumed to be twice its short-circuit inductance. Table~\ref{tab:DTs} summarizes the parameters for different DT sizes.
\begin{table}[htbp]
\centering
\caption{Distribution Transformer Parameters}\label{tab:DTs}
\begin{tabular}{c c c c c c c}
\toprule
Capacity (kVA) & 25 & 50 & 75 & 100 & 125 & 150\\
\midrule
Voltage Drop (\%) & 1.0 & 1.5 & 2.0 & 2.5 & 3.2 & 4.0\\
Load Losses (W) & 201 & 340 & 469 & 571 & 706 & 827\\
\bottomrule
\end{tabular}
\end{table}
\subsection{Security Limits for Frequency-related
Metrics}\label{frequencylimits}
 The system's operational security limits for frequency-related metrics are defined as follows: frequency and QSS frequency ranges of $(59.50\sim 60.50)$ Hz, frequency nadir range of $(57.80\sim 61.80)$ Hz, and maximum RoC of frequency range of $(-4.00\sim 4.00)$ Hz/s.
\ifCLASSOPTIONcaptionsoff
  \newpage
\fi
\bibliographystyle{IEEEtran}
\bibliography{IEEEabrv, references.bib}
\end{document}